TITLE

# Near-Room Temperature Ferromagnetic Insulating State in Highly Distorted LaCoO$_{2.5}$ with CoO$_5$ Square Pyramids

**AUTHORS**

Qinghua Zhang[1,2], Ang Gao[1,3], Fanqi Meng[1,3], Qiao Jin[1,3], Shan Lin[1,3], Xuefeng Wang[1], Dongdong Xiao[1], Can Wang[1,3,4], Kui-juan Jin[1,3,4], Dong Su[1], Er-Jia Guo[1, 4, 5, *], and Lin Gu[1, 3, 4, *]

[1] Beijing National Laboratory for Condensed Matter Physics, Institute of Physics, Chinese Academy of Sciences, Beijing 100190, China

[2] Yangtze River Delta Physics Research Center Co. Ltd., Liyang 213300, China

[3] School of Physical Sciences, University of Chinese Academy of Sciences, Beijing 100049, China

[4] Songshan Lake Materials Laboratory, Dongguan, Guangdong 523808, China

[5] Center of Materials Science and Optoelectronics Engineering, University of Chinese Academy of Sciences, Beijing 100049, China

*E-mail: ejguo@iphy.ac.cn; l.gu@iphy.ac.cn

**ORCID**:

**Dong Su**: 0000-0002-1921-6683

**Er-Jia Guo**: 0000-0001-5702-225X

**Lin Gu**: 0000-0002-7504-031X

**ABSTRACT**

Dedicated control of oxygen vacancies is an important route to functionalizing complex oxide films. It is well-known that tensile strain significantly lowers the oxygen vacancy formation energy, whereas compressive strain plays a minor role. Thus, atomically reconstruction by extracting oxygen from a compressive-strained film is challenging. Here we report an unexpected $LaCoO_{2.5}$ phase with a zigzag-like oxygen vacancy ordering through annealing a compressive-strained $LaCoO_3$ in vacuum. The synergetic tilt and distortion of $CoO_5$ square pyramids with large La and Co shifts are quantified using scanning transmission electron microscopy. The large in-plane expansion of $CoO_5$ square pyramids weaken the crystal-field splitting and facilitated the ordered high-spin state of $Co^{2+}$, which produces an insulating ferromagnetic state with a Curie temperature of ~284 K and a saturation magnetization of ~0.25 $\mu_B$/Co. These results demonstrate that extracting targeted oxygen from a compressive-strained oxide provides an opportunity for creating unexpected crystal structures and novel functionalities.

**INTRODUCTION**

Dedicated control of oxygen defects has been considered an important route to functionalizing complex oxide thin films. The nature of oxygen vacancies (Vo) in thin films has been utilized in the redox chemistry and energy applications[1-4]. Recently, it has been recognized that the formation and thermodynamically motion of Vo in oxide thin films can be effectively controlled by epitaxial strain. Oxygen vacancies (Vo) related fascinating properties by utilizing epitaxial tensile strain have been widely reported such as in the nickelates[5], manganites[6], and ferrites[7]. Tensile strain would lower the Vo formation energy in most oxides, in favor of a high Vo concentration[8-10]. It can create Vo even in a highly oxidizing environment, resulting in the enhancement of catalytic activity by up to an order of magnitude[11]. However, strain effects on the Vo formation are nonlinear but highly anisotropic; the compressive strain would not play a significant role in engineering the oxygen content of oxide thin films. As a matter of fact, it strongly limits the possibility of tailoring the physical properties by manipulating their oxygen stoichiometry in a compressively strained film. Many fascinating oxides have had their performance enhanced or have been enticed into a ferroic ground state with compressive strain. For instance, the record-high ferroelectric polarization up to ~ 130 μC/cm$^2$ achieved in the tetragonal-like BiFeO$_3$ can be stabilized under an extremely large compressive strain greater than −4.5%[12]. Compressively strained EuTiO$_3$ with enhanced spin-lattice coupling exhibits emergent multiferroicity with the strongest ferroelectric and ferromagnetic properties known today[13-15]. Therefore, understanding how the kinetic behavior of oxygen loss affects the structure of compressively strained oxides is critical for identifying the true nature of physical properties in a multitude of functional oxide films.

An excellent example of strain-mediated physical property can be found in studies of lanthanum cobaltite thin films, LaCoO$_3$ (LCO$_3$), in which the active spin-state transition and intriguing magnetic ground state vary dramatically with lattice distortion[16-20]. The spin state of Co ions is known to be reversibly controlled by the delicate competition between crystal field splitting and Hund's exchange coupling[16]. Tensile-strained LCO$_3$ films exhibit an insulating behavior and a robust ferromagnetic ordering[17-20], suggesting potential applications toward low-power spintronic devices (e.g., spin filters and spin-based logics), in which the spin current carries the encoded information without mobile charges[21, 22]. However, the Curie temperature ($T_C$) of tensile-strained LCO$_3$ films is independent as tensile strain increases and keeps a constant value of ~80 K, limiting their practical applications. Thus, there is an urgent need to increase $T_C$ of LCO in order to achieve high-temperature FMI. In the case of LCO films under compressive strain, a general consensus is that the exchange between different Co ions is suppressed by compression and restrains the long-range magnetic ordering. Meanwhile, the stoichiometry and lattice structure of a compressively strained LCO film does not change at ambient conditions.

Here we report a near-room temperature ferromagnetic insulating (FMI) phase stabilized in lanthanum cobaltite thin films with previously unknown lattice structure. The metastable phase with stoichiometry of LaCoO$_{2.5}$ is naturally formed through vacuum annealing by extracting oxygen from a compressively strained LCO$_3$ film.

Compared with the antiferromagnetic brownmillerite (BM) LaCoO$_{2.5}$, this unexplored LaCoO$_{2.5}$ phase exhibits an ideal FMI behavior with $T_C$ of ~284 K and a saturation magnetization of ~0.25 μ$_B$/Co. It exhibits a zigzag-like Vo ordering and consists of CoO$_5$ square pyramids with large atomic shifts of both La and Co ions. The unique Vo pattern mediated by compressive stress induces a large in-plane expansion of CoO$_5$ square pyramids, which weakens crystal field splitting and leads to the high-spin filling of $d$ orbitals, and produces the FMI state close to room temperature, which agrees with our first-principles calculations. We believe that the similar emergent phase mediated by compressive strain may also stimulate further studies on many other multifunctional oxides, such as nickelates and ferrites with relatively low oxygen vacancy formation energy.

## RESULTS

**Vo orderings evolution and emergent near-room temperature ferromagnetism.**

High-quality stoichiometric LCO$_3$ films with a thickness of 200 unit cells (u.c.) were grown on (001)-oriented LaAlO$_3$ (LAO) substrates using pulsed laser deposition (see Methods). The LCO$_3$ layer was subsequently capped with a 50-u.c.-thick SrTiO$_3$ layer to prevent the intrinsic non-stoichiometry. We first investigated the microstructure of the LCO$_3$ films on (001)-oriented LaAlO$_3$ (LAO) substrates using scanning transmission electron microscopy (STEM) and selected area electron diffraction (SAED). The cross-sectional STEM measurements illustrate both high crystallinity and an atomically sharp interface between the film and substrate (Figure 1a). The SAED pattern in Figure 1a confirms that LCO$_3$ films are coherently strained. Structural inhomogeneities, such as dark stripes or Vo stripes[19, 23], were not detected in pristine LCO$_3$, which indicates that the film was nearly stoichiometric without visible oxygen vacancies. We performed *in-situ* sample-annealing. After annealed 2 hours in vacuum, the dark stripes in every third Co (001) plane with a diffraction vector of 1/3 (0, 0, 1) appear in the film, as labeled by small yellow arrows in Figure 1b. This phase is recognized as LaCoO$_{2.67}$ (LCO$_{2.67}$) or La$_3$Co$_3$O$_8$ [24], which is in agreement with previous reports that indicated that compressive-strained LCO$_{2.67}$ favored in-plane Vo stripes[25]. We continuously annealed the sample in vacuum for 4 hours. Surprisingly, a previously unknown ordered dark stripe pattern with Vo ordering is aligned along the diagonal direction, i. e., [011] orientation appears, as shown in Figure 1c. The SAED pattern of this structure possesses a diffraction vector of 1/3 (0, 1, 1). Compared to earlier works, the Vo ordering and SAED pattern are completely different from the brownmillerite-type (BM) LCO$_{2.5}$ structure, in which the alternative tetrahedral CoO$_4$ layers and octahedral CoO$_6$ layers are stacked and a diffraction vector along the 1/2 (0, 0, 1)[26]. To our best knowledge, this type of Vo ordering pattern has never been reported in any oxygen-deficient perovskite oxides[26]. High magnification HAADF-STEM images and corresponding FFT patterns of these structures are summarized in Supplementary Fig. 1 for a close comparison. Both samples exhibit atomically sharp interfaces and highly coherent epitaxial films after the thermal processes (Supplementary Fig. 2).

We performed EELS measurements of both O *K*-edge and Co *L*-edges to quantify the oxygen concentration of these structures (Supplementary Fig. 3). The significant decrease of pre-peak's intensity at the O *K* edge and concomitant chemical shift at Co $L$ edges suggest that the number of Vo increases with doubling of the annealing time. Quantified analysis of the O/Co atomic ratio by the standard procedure as implemented in DigitalMicrograph$^{TM}$ allows us to estimate the oxygen content of ~ 2.70 $\pm$ 0.07 and ~ 2.53 $\pm$ 0.07 for LCO$_{2.67}$ and LCO$_{2.5}$ phase, respectively. To differentiate it from the BM LCO$_{2.5}$ phase, this unexplored phase is hereafter referred to as nLCO$_{2.5}$. Macroscopic structure and topography characterizations confirm the highly epitaxial, coherent growth, smooth surface, and single-phase of pristine LCO$_3$ and nLCO$_{2.5}$ films (Supplementary Fig. 4). X-ray absorption spectroscopy measurements demonstrate that the Co$^{3+}$ ions in LCO$_3$ transit into Co$^{2+}$ ions in nLCO$_{2.5}$ due to the formation of Vo. Transport measurements demonstrate that both LCO$_3$ and nLCO$_{2.5}$ films exhibit insulating behavior (Supplementary Fig. 5).

The magnetic properties were examined in LCO$_x$ samples with different annealing times. The temperature ($T$)- and magnetic field ($H$)-dependent magnetization ($M$) of pristine LCO$_3$ and nLCO$_{2.5}$ samples, respectively, is shown in Figures 1d-1f. Pristine LCO$_3$ film exhibits the typical diamagnetic or paramagnetic behavior, which is in agreement with earlier reports[27]. However, nLCO$_{2.5}$ exhibits a clear magnetic phase transition at the Curie temperature ($T_C$) of ~284 K and square-like hysteresis loops, which indicates a ferromagnetic character in nLCO$_{2.5}$. At 10 K, the continuous increase of magnetization with increasing magnetic field suggests a small portion of paramagnetic component in the nLCO$_{2.5}$, consistent with a large upturn in $M$-$T$ curve at low temperatures. The saturation magnetic moment ($M_S$) of nLCO$_{2.5}$ reaches ~35 emu/cm$^3$ (~0.25 $\mu_B$/Co) at 200 K. Furthermore, we also measured the magnetic properties of $LCO_{2.67}$. $T_C$ of $LCO_{2.67}$ is similar to that of $nLCO_{2.5}$, but $M_S$ of $LCO_{2.67}$ is an order of magnitude smaller than that of $nLCO_{2.5}$ (Supplementary Fig. 6). We believe that the small magnetization in $LCO_{2.67}$ may be attributed to the presence of small amount $nLCO_{2.5}$ phase in $LCO_{2.67}$ films.

**Atomic-scale lattice and EELS analysis of two different Vo orderings**

The atomistic details of $LCO_{2.67}$ and $nLCO_{2.5}$ with two different Vo orderings are examined via atomic-resolved HAADF images and EELS spectra. The HAADF image of the $LaCoO_{2.67}$ lattice with Vo stripes in every third Co (001) plane is shown in Figure 2a, which is also manifested by the periodic chemical expansion (~4.53 Å) of the out-of-plane La–La distances (Figure 2b) and the periodic intensity decrease of O $K$-edges (Figures 2c, 2d and Supplementary Fig. 7) in the oxygen-deficient tetrahedral $CoO_4$ layer. We extract the fine structures of the O $K$-edge and Co $L$-edges in both tetrahedral and octahedral layers, respectively. A clear broadness of the main peak in the O $K$-edge and a narrowing of Co $L_3$- and $L_2$-edges in the tetrahedral $CoO_4$ layer can be clearly visualized. Furthermore, a chemical shift toward lower energy and increased Co $L_3/L_2$ ratio in the tetrahedral $CoO_4$ layer also indicates the reduced valence state of Co ions[28]. These results provide atomic-scale evidence for the alternative stacking of one tetrahedral $CoO_4$ layer (yellow polyhedral) with two octahedral $CoO_6$ layers (green polyhedral), which is consistent with a previous report[24]. In contrast, a highly distorted lattice with Vo stripes in every third O $(011)_p$ plane appears in the $nLCO_{2.5}$ film (Figure 2g), which is characterized by alternative chemical expansion (~4.60 Å) between out-of-plane La–La distances (Figure 2h), resulting in the concomitant in-plane wave-like atomic arrangement of La. The respective upward and downward shifts of two La that are vertically close to Vo sites lead to the dark contrast; the ordering of which produces dark stripes along the (011) plane (Figure 2g). Meanwhile, the periodic intensity decreases of O $K$ edges (Figures 2i and 2j) confirms Vo sites. Similar peak positions and intensity distribution in the fine structures of Co $L$ edges on each Co sites are observed (Supplementary Fig. 8), which suggests the same type of [CoO$_x$] polyhedra. Combining with the EELS quantification analysis shown in Supplementary Fig. 3, the nLCO$_{2.5}$ phase should consist of all [CoO$_5$] square pyramids.

**Quantitative analysis of atomic shifts and tilt of CoO$_5$ square pyramids**

To determine the exact structure of the nLCO$_{2.5}$ phase, we used a recently developed integrated differential phase contrast (iDPC) imaging technique [29-31] to identify atomic positions of oxygen and Vo. The iDPC-STEM enables the linear imaging of the projected electrostatic potential of atomic columns, resulting in contrast mechanism nearly proportional to the atomic number $Z$ instead of its square in the HAADF-STEM. This method is extremely sensitive to light elements (e.g., O) because of the linear phase contrast mechanism, which provides better signal-to-noise ratio and enhanced accuracy of atomic positions [30, 31]. As shown in Figure 3a and Supplementary Fig. 9, the wave-like atomic shifts of La (big green circles), cooperative displacement of O (small red spheres), and Co (pink spheres) can be directly identified. The remarkable absence of image contrast in some oxygen sites manifests itself as Vo rows along the viewing direction. By overlaying pink polygon, the tilt and distortion of CoO$_5$ square pyramids can be clearly visualized (Figure 3b), i.e., the clockwise rotation in the projected triangles of [Co1–O5] and [Co1′–O5] pyramids close to Vo sites and the counterclockwise rotation in the projected quadrilateral of [Co2–O5] pyramids.

We conduct quantitative analysis of the lattice distortion and $CoO_5$ square pyramid tilting on the iDPC images by extracting the atomic positions of La, Co, and O using the Calatom software[32]. The vertical La shifts caused by Vo, with the magnitude of ~0.3 Å, are identified in Figure 3c. The in-plane projected Co–Co distance exhibits a distinctive reduction in every third unit cell, decreasing from 3.90 to 3.49 Å. This reduction is unusual because the atomic distances across Vo sites always increase because of chemical expansion, e.g., La–La distances, as shown in Figures 2a, 2g, and 3a. Most likely the abnormal contraction of the in-plane Co–Co distance is the result of the interplay between this zigzag-like Vo and compressive strain state, which leads to the highly elongated lattice along the out-of-plane direction for better misfit accommodation. Besides the large displacement of La and Co ions, oxygen positions are also quantitatively extracted, which allows to perform accurate statistics on $CoO_5$ square pyramids. As shown in Figure 3e, the statistical rotation of [Co1–O5], [Co2–O5], and [Co1′–O5] square pyramids are $6.0 \pm 0.6°$, $-17.4 \pm 0.8°$, and $15.8 \pm 1.1°$, respectively, which considerably changes the in-plane Co–O–Co bond angle.

**DISCUSSION**

To further clarify the origin of ferromagnetic insulating properties, we employ first-principles calculations on the $nLCO_{2.5}$ phase using the Heyd–Scuseria–Ernzerhof (HSE06) hybrid functional. According to the distribution of Vo from STEM, a $\sqrt{5} \times 2\sqrt{2} \times 2$ supercell is adopted (Supplementary Fig. 10), which is optimized with fixed lattice parameters based on the LaAlO$_3$ substrate (Figure 4a and Supplementary Table 1). Three distinct types of [Co–O5] pyramids, named [Co1–O5], [Co1′–O5], and [Co2–O5], were identified with different Co–O coordination, which are distributed in the nearest-neighbor and next nearest-neighbor sites of the Vo channels. The Co–O bonds in the subface (*xy* plane) of [Co2–O5] pyramids clearly expand compared with [Co1–O5] and [Co1′–O5] pyramids, as shown in Figure 4b and Supplementary Table 2. The electrostatic field of [CoO5] square pyramids cause 3*d* orbitals to split into a doubly degenerate pair ($d_{xz}$, $d_{yz}$) and three singly degenerate $d_{xy}$, $d_{z^2}$, and $d_{x^2-y^2}$ levels[33]. As shown in Figure 4, the higher $d_{x^2-y^2}$ level due to short Co–O bonds in the *xy* plane accounts for the low-spin state in [Co1–O5] and [Co1′–O5] pyramids. By contrast, the expansion in the *xy* plane of [Co2–O5] pyramids result in a weaker crystal field splitting and lowers the $d_{x^2-y^2}$ level, which leads to electron filling and the high-spin state of Co2 ions.

The Density of states (DOS) is used to elucidate spin states of Co1, Co1′, and Co2 originating from distorted crystalline field, where the obvious asymmetric spin occupation of Co2 promises a high-spin state (Figure 4c). The integrated spin of Co1, Co1′, and Co2 is −0.77, −0.91, and 2.54 μ$_B$, respectively, as shown in Figure 4d. It produces the total net magnetic moment determined by the following equation:

$$\frac{(2.54\mu_B \times 4 - 0.77\mu_B \times 4 - 0.91\mu_B \times 4)}{12} = 0.29\ \mu_B/Co$$

which is in good agreement with experimental magnetic moment (0.25 μ$_B$/Co).

The insulating nature of the nLCO$_{2.5}$ phase is attributed to the ordering of Vo[19], which is confirmed by the total DOS result. The occupied states closer to the Fermi energy is primarily composed of O-*p* and Co-*d* states. There exists a slightly different DOS distribution between Co1 and Co1′, as shown in Figure 4c, which may be attributed to the different rotation magnitude of $CoO_5$ square pyramids, as shown in the iDPC image (Figure 3e). Except for a small contribution near the Fermi level from the high-spin state of Co1, a bandgap of 1.1 eV is obtained, as displayed in Figures 4c and 4e. Thus, oxygen ordering and compressive strain work synergistically to lower the crystal-field splitting energy and produce the FMI behavior.

In summary, we report a high-temperature FMI state in a previously unknown lattice structure of $LaCoO_{2.5}$ films by gently extracting oxygen from a compressively strained $LaCoO_3$ film. Our result challenges the present hypothesis on the indispensability of tensile stress in the origin of ferromagnetism in $LaCoO_3$ films. More importantly, the compressive-strain-induced peculiar Vo ordering causes synergetic tilt and distortion of $CoO_5$ square pyramids, where the large in-plane expansion weakened the crystal-field splitting and facilitated the ordered high-spin state of $Co^{2+}$ and an insulating ferromagnetic state near room temperature. These results suggest that compressive stress can be reconsidered as being important for the reconstruction of Vo ordering pattern in other oxygen-deficient functional oxide films, and emergent crystal structures and promising functionalities can be expected even in the non-perovskite structural families that are relevant include chrysoberyl[34], pyrochlore[35], and delafossite[36].

**METHODS**

**Sample preparation.** Pulsed laser deposition (PLD) was used to fabricate single-crystalline LCO films by ablating a stoichiometric ceramic target. The LCO films with a thickness of 200 u.c. were grown on LAO substrates at a substrate temperature of 700 ºC and under an oxygen partial pressure of 100 mTorr. During the deposition, the laser frequency and energy density were kept as 5 Hz and 1.5 J/cm$^2$, respectively. A 50-u.c.-thick SrTiO$_3$ layer was subsequently capped on top of LCO layer to prevent formation of oxygen vacancies at the surfaces. After the film growth, the pristine LCO$_3$ samples were cooled down to room temperature in an oxygen environment of 100 Torr to avoid oxygen vacancies. The LCO$_{2.67}$ and nLCO$_{2.5}$ samples were formed by annealing in vacuum ($P \sim 1\times10^{-7}$ Torr) at 600 ºC for two and four hours, respectively. X-ray reflectivity, diffraction measurements, and reciprocal space mapping were conducted on D8 Discovery diffractometer. The thickness of each layer was calibrated by x-ray reflectivity measurements. The magnetic properties of all samples were probed using SQUID. *M-T* curves were recorded during the warming up under a small field of 0.1 T after the samples were either zero-field-cooled (ZFC) or field-cooled (FC).

**STEM characterizations.** Sample was prepared by using focused ion beam (FIB) milling. Cross-sectional lamellas were thinned down to 100 nm thick at an accelerating voltage of 30 kV with a decreasing current from the maximum 2.5 nA, followed by fine polish at an accelerating voltage of 2 kV with a small current of 40 pA. The atomic structures of the LCO, LCO$_{2.67}$ and nLCO$_{2.5}$ films was characterized using an ARM 200CF (JEOL, Tokyo, Japan) transmission electron microscope operated at 200 kV and equipped with double spherical aberration (Cs) correctors. HAADF images were acquired at acceptance angle of 90~370 mrad. The iDPC-STEM imaging was conducted using a Cs-corrected (S)TEM (FEI Titan Cubed Themis G2 300) with a convergence semi-angle of 15 mrad, operated at a voltage of 300 kV. The collection angle for the iDPC-STEM imaging is 4~20 mrad. The STEM was equipped with a DCOR+ spherical aberration corrector for the electron probe which was aligned using a standard gold sample before observations. Four images used for 2D integration were acquired by a 4-quadrant DF4 detector with an optional high-pass filter applied to reduce the low frequency information in the image.

**Calculation details.** All first-principles calculations were performed within the Vienna *Ab Initio* Simulation Package (VASP) based on the density functional theory (DFT). The projected augmented wave (PAW) potentials were used to deal with the electronic exchange-correlation interaction along with GGA functional in the parameterization of Perdew Burke and Ernzerhof (PBE) pseudopotential. A plane wave representation for the wave function with a cut off energy of 500 eV was applied. Geometry optimizations were performed using a conjugate gradient minimization until all the forces acting on ions were less than 0.01 eV/Å per atom. The K-point mesh with a spacing of ca. 0.03 Å$^{-1}$ was adopted. Crystal structures are built using VESTA software. The in-plane lattice parameters were fixed to the LaAlO$_3$ substrate (3.79 Å), as experimental proved by x-ray diffraction measurements. The La$_{12}$Co$_{12}$O$_{30}$ structure with a supercell of √5×2√2×2 was adopted for the structural simulation (Supplementary Fig. 7a). All calculations of electronic structure are based on the HSE06 hybrid functional with hybrid mixing parameters ($\alpha$=0.25).

## DATA AVAILABILITY

The data that support the findings of this study are available on request from the first author (Q.H.Z.) and the corresponding authors (E.J.G and L.G.).

**ACKNOWLEDGEMENTS**

This work was supported by the National Key Basic Research Program of China (Grant Nos. 2020YFA0309100 and 2019YFA0308500), Beijing Natural Science Foundation (Z190010 and 2202060), the Strategic Priority Research Program of Chinese Academy of Sciences (Grant Nos. XDB07030200 and XDB33030200), the National Natural Science Foundation of China (Grant Nos. 51672307, 51991344, 11974390, 52025025, 52072400), the Beijing Nova Program of Science and Technology (Grant No. Z191100001119112). XAS experiments were conducted at the beamline 4B9B of Beijing Synchrotron Radiation Facility (BSRF) of the Institute of High Energy Physics, Chinese Academy of Sciences were conducted via user proposals.

**AUTHOR CONTRIBUTIONS**

These samples were grown and processed by Q. J. under the guidance of E.J.G. and K.J.J; TEM lamellas were fabricated with FIB milling by F.Q.M and D.D.X.; TEM experiments were performed by Q.H.Z. and F.Q.M, analyzed by Q.H.Z., X.F.W. and D.S; The first-principles calculations were performed by A.G. Q.J. and C.W. worked on the magnetic measurement. Q.H.Z. and E.J.G. initiated the research and L.G. supervised the work. All authors participated in writing the manuscript.

**COMPETING INTERESTS**

The authors declare no competing interests.

**ADDITIONAL INFORMATION**

**Supplementary information** accompanies this paper at http://......

**Reprints and permissions information** is available online at http://npg.nature.com/reprintsandpermissions/

**Journal peer review information**: Nature Communications thanks the anonymous reviewers for their contribution to the peer review of this work.

**Publisher's note**: Springer Nature remains neutral with regard to jurisdictional claims in published maps and institutional affiliations.

**Correspondence and requests for materials** should be addressed to E.J.G. and L.G.

**FIGURES AND FIGURE CAPTIONS**

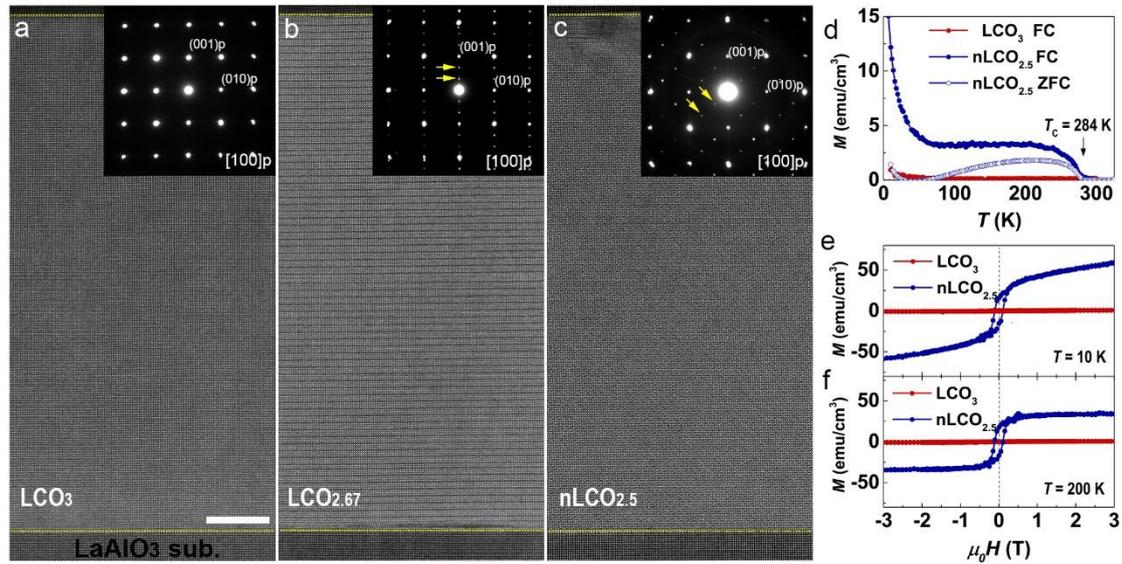

**Figure 1. Structural evolution and magnetic properties of LCO films during anneal process.** Low magnification high-angle angular dark-field (HAADF) images of **a** pristine $LCO_3$, **b** $LCO_{2.67}$, and **c** $nLCO_{2.5}$ films grown on LAO substrates. The corresponding selected-area electron diffraction (SAED) patterns are inserted on the right top corner of each panel, where the yellow arrows indicate the positions of superstructure spots. The white bar represents 10 nm in scale. **d** *M-T* curves of $LCO_3$ and $nLCO_{2.5}$ films. The measurements were carried out during the sample warm-up under a magnetic field of 0.1 T. Solid and open symbols represent the field-cooled (FC) and zero-field-cooled (ZFC) data, respectively. **e** and **f** *M-H* curves of $LCO_3$ and $nLCO_{2.5}$ films at 10 K and 200 K, respectively.

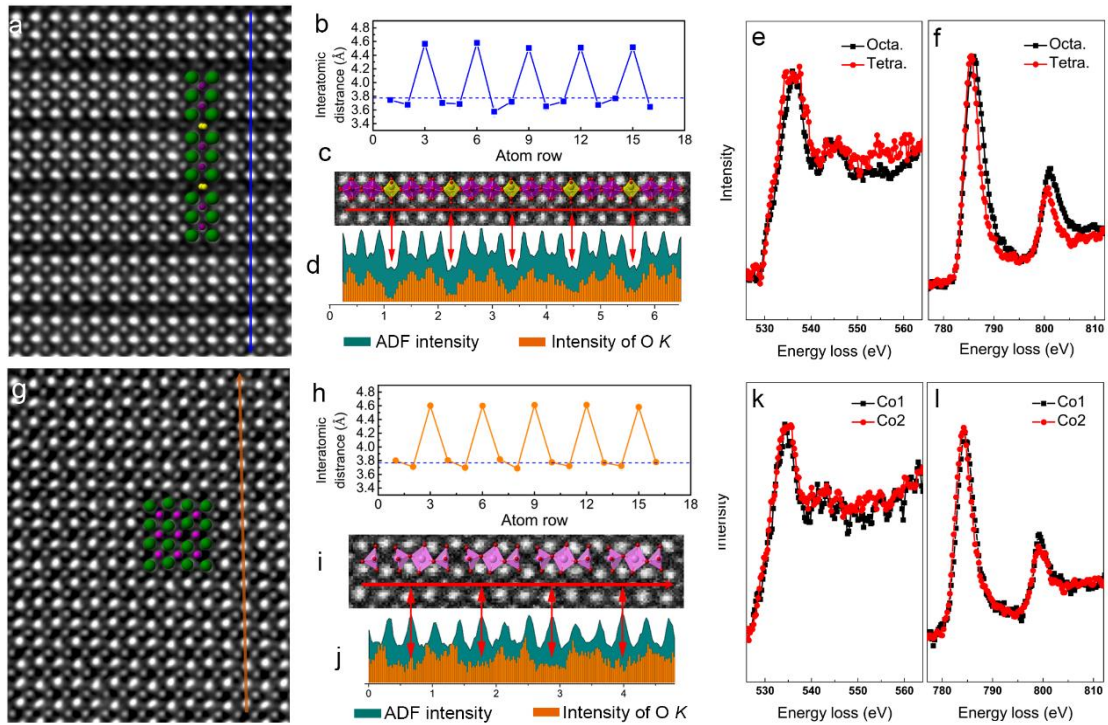

**Figure 2. Atomic-resolved HAADF images and electron energy loss spectra of LCO$_{2.67}$ and nLCO$_{2.5}$ phases.** HAADF images of **a** LCO$_{2.67}$ and **g** nLCO$_{2.5}$ phase with overlaid structural model, where green spheres represent La, purple and yellow spheres represent Co in octahedral and tetrahedral layers, the O are omitted for clarity. Dark blue and yellow lines indicated the acquired positions of EELS spectra. Periodic change of out-of-plane La-La distances in the **b** LCO$_{2.67}$ phase and **h** nLCO$_{2.5}$ phase, the 3.79 Å is indicate by dotted line. Local HAADF images overlaid with projected ployhedra and corresponding periodic change of intensity of ADF signal and O K edge in LCO$_{2.67}$ phase (**c, d**) and nLCO$_{2.5}$ phase (**i, j**). Red double arrows indicate the correspondence between the HAADF image and EELS results. The normalized O *K*-edge and Co *L*-edges of the LCO$_{2.67}$ phase (**e, f**) and the nLCO$_{2.5}$ phase (**k, l**).

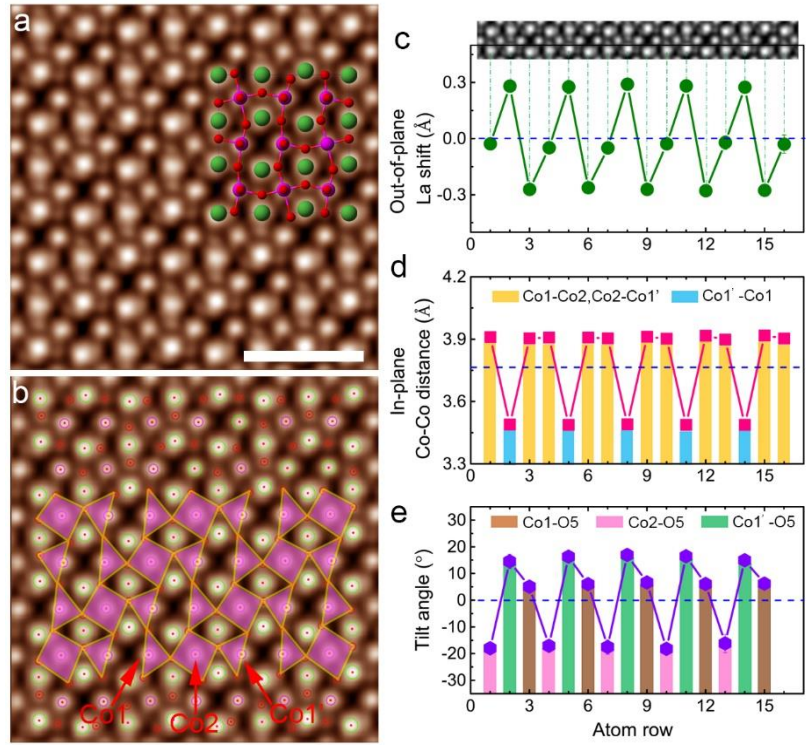

**Figure 3. Quantitative analysis on the atomic shift of La, Co ions, and tilt of [CoO5] square pyramids. a** The iDPC image of the nLCO$_{2.5}$ phase. The structural model is overlaid to show the contrast assignment of La (green), Co (pink) and O (red) atoms. The scale bar is 1 nm. **b** Overlaid iDPC image in **a** by circles and polyhedra, where La, Co and O atom columns are outlined by blue, pink and red circles, respectively; projected CoO$_5$ square pyramids are labeled by pink polyhedral; three kinds of Co sites are also indicated by red arrows. **c** Periodic out-of-plane atomic shifts of La, where the HAADF image is inserted on the top to demonstrate the correspondence relationship. **d** Periodic in-plane Co-Co distances, where relatively small Co1'-Co1 and large Co1-Co2, Co2-Co1' distances are indicated by blue and yellow columns, respectively. **e** Tilt angles of CoO$_5$ square pyramids on Co1, Co2 and Co1' sites, which are indicated by brown, pink and green columns, respectively.

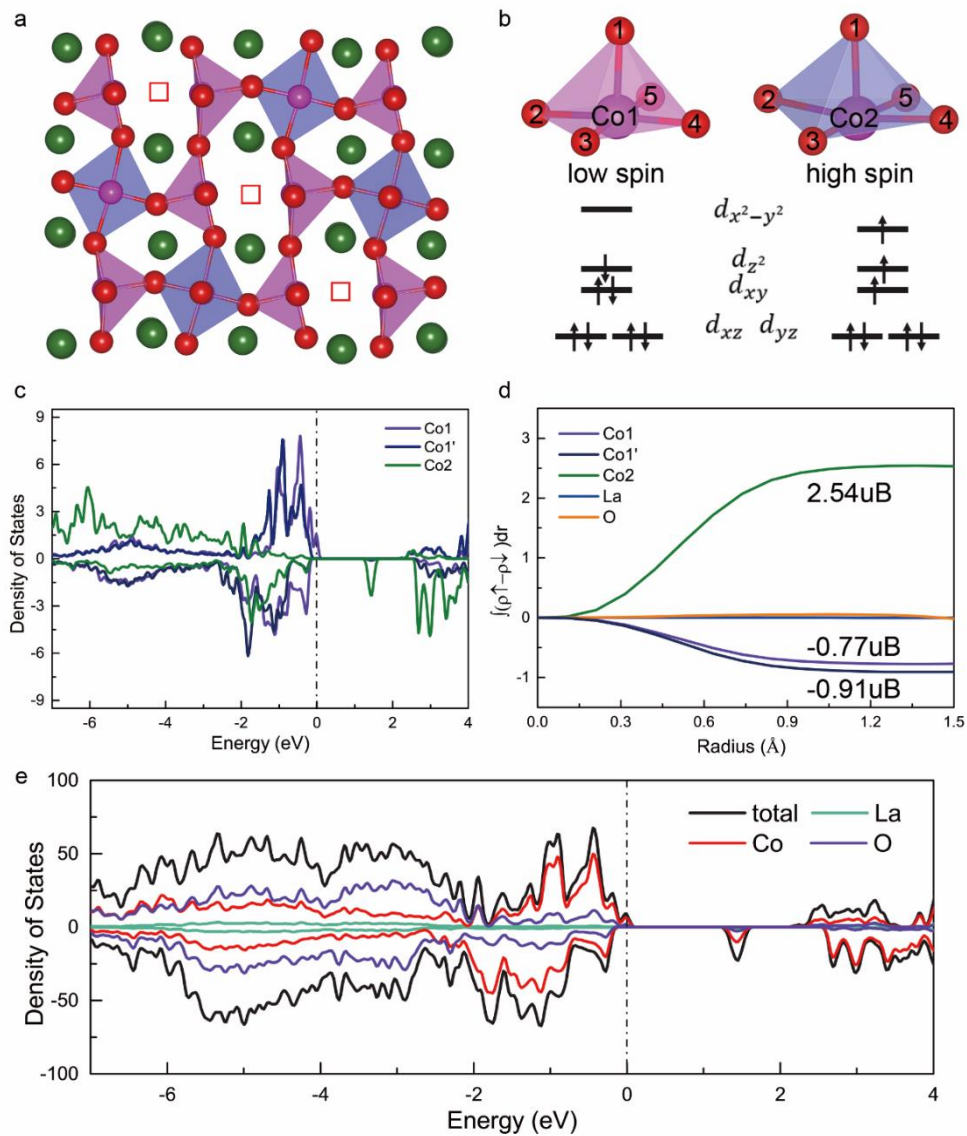

**Figure 4. The atomic and electronic structure of the** nLCO$_{2.5}$ **phase. a** The optimized structures of nLCO$_{2.5}$ phase. **b** Schematic diagram of Co 3*d* orbits splitting in [Co1-O5] and [Co2-O5] tetragonal pyramid, including the low spin of Co1 and high spin of Co2. **c** Projected density of states of Co1, Co1' and Co2. **d** Integrated spin as a function of the radius around Co1, Co1', Co2, La and O ions. **e** Projected density of states of nLCO$_{2.5}$ structure.